\documentclass[12pt]{article} 
\usepackage{epsfig}  
\def\tozero#1
{\mathrel{\mathop{\to}\limits_{\scriptscriptstyle {#1\rightarrow 0 }}}}
\def\bea{\begin{eqnarray}}
\def\eea{\end{eqnarray}}
\def\beq{\begin{equation}}  
\def\eeq{\end{equation}}  
\def\as{\alpha_s}

\def\abs#1{\left|#1\right|}

\newenvironment{comment}[1]{}{}
\newcommand{\ls}{\ell\hspace{-1.2ex}/}
\newcommand{\ps}{p\hspace{-1.2ex}/}

\newcommand{\pps}{p'\hspace{-1.6ex}/}

\newcommand{\lamb}{\lambda_b}
\newcommand{\lamc}{\lambda_c}
\newcommand{\non}{\nonumber}
\newcommand{\mg}{\lambda}
\newcommand{\hmg}{\hat \lambda}
\newcommand{\hQt}{\hat\omega}
\newcommand{\Qt}{\omega}
\newcommand{\hQf}{\hat\omega^2}
\newcommand{\Qf}{\omega^2}

\newcommand{\hmct}{\rho}
\newcommand{\hmc}{\sqrt{\rho}}

\catcode`\@=11
\@addtoreset{equation}{section}

\begin{document}
\pagestyle{empty}
\begin{flushright}   
  
Bicocca-FT-05-5\\DFTT/06-05\\GeF/TH/1-05\\LMU-ASC 20/05\\

\end{flushright}   
  
\begin{center}   
\vspace*{0.5cm}  
{\LARGE \bf Perturbative corrections to semileptonic\\
  \vspace*{3mm}  \boldmath $b$ decay distributions} \\   
\vspace*{1.5cm}   
V. Aquila$^{a}$, P. Gambino$^{b}$,
G. Ridolfi$^{c}$ and N. Uraltsev$^{d\, *}$\\  
\vspace{0.6cm}  
{\it
{}$^a$Arnold Sommerfeld Center, Department f\"ur Physik,\\
Ludwig-Maximilians-Univ.\ M\"unchen, Theresienstra{\ss}e 37,
D-80333 M\"unchen, Germany\\ 
\medskip
{}$^b$INFN, Sezione di Torino, and Dipartimento di Fisica Teorica, 
Universit\`a di Torino\\  
Via P. Giuria 1, I-10125 Torino, Italy\\  
\medskip
{}$^c$INFN, Sezione di Genova, and Dipartimento di Fisica, 
Universit\`a di Genova\\    
Via Dodecaneso 33, I-16146 Genova, Italy\\  
\medskip
{}$^d$INFN, Sezione di Milano, Italy and\\
Department of Physics, University of Notre Dame du Lac,
Notre Dame, IN 46556, USA}
\vspace*{1.5cm}  

{\bf Abstract}  
\end{center}  
  
\noindent  
We compute $O(\alpha_s)$ and $O(\alpha_s^n \beta_0^{n-1})$ (BLM)
corrections to the five structure functions relevant for $b\to q \ell
\bar\nu$ decays and apply the results to the moments of a few
distributions of phenomenological importance.  We present compact
analytic one-loop formulae for the structure functions, with proper
subtraction of the soft divergence.

\vspace*{1cm}  

\vfill  
  ~\hspace*{-12.5mm}\hrulefill \hspace*{-1.2mm} \\
\footnotesize{
\hspace*{-5mm}$^*$
On leave of absence from
St.\,Petersburg Nuclear Physics
Institute, Gatchina, St.\,Petersburg 188300, Russia}
\normalsize
\eject   

\pagestyle{plain}
\section{Introduction}

The experimental study of inclusive semileptonic $B$ decays has
reached an unprecedented level of precision.  The total branching
ratio is now known to better than 2\% \cite{babar}, while various
moments of the lepton energy and invariant hadronic mass distributions
have been measured by the Delphi, BaBar, Belle, CLEO, and CDF
Collaborations \cite{moments_exp} with impressive accuracy.  The first
field of application of this wealth of data is the extraction of the
CKM matrix element $V_{cb}$, but the semileptonic distributions
provide us with important information on the structure of the B meson
and on the quark masses that can be employed in many applications.

From the theoretical point of view, inclusive semileptonic $B$ decays
can be described in the framework of the Operator Product Expansion
(OPE) in inverse powers of the heavy $b$ mass. The dominant term in
this expansion represents the decay of a free bottom quark and the
non-perturbative corrections to the parton model result are suppressed
by at least two powers of the energy release, $E_{rel}\approx m_b-m_c$
\cite{1mb2}.  The $O(1/m_b^2)$ \cite{1mb2,1mb2bis} and $O(1/m_b^3)$
\cite{1mb3} corrections are known in terms of the $B$ meson
expectation values of a limited set of dimension five and six
operators.  These non-perturbative parameters can be extracted from
the moments of the measured spectra of semileptonic and radiative
inclusive decays, although complementary information is also provided
by a number of heavy quark relations and sum rules (see {\it e.g.}
\cite{Uraltsev:2001ih}).  Recent fits to the moments of the
distributions lead to consistent results and allow for a determination
of $|V_{cb}|$ to better than 2\% \cite{babar,delphi,Bauer:2004ve},
with theoretical uncertainties starting to play a dominant role
\cite{imprecated}.

At the current level of experimental accuracy it becomes crucial to
have perturbative corrections under control.  In this respect, a lot
of work has been done in the past 25 years.  The one-loop corrections
to a few differential distributions have been known for a long time
\cite{Ali:1979is,jezabek} and the correction to the total width is
known in terms of polylogarithms since 1989 \cite{Nir:1989rm}.
Analytic expressions for the $O(\alpha_s)$ corrections to the moments
of the hadron spectra have been computed in \cite{fls95,bauer}, while
numerical results for the leptonic and hadronic moments with a lower
cut on the charged lepton energy can be found in \cite{voloshin} and
\cite{flcut,bauer}.  However, the complete one-loop corrections to the five
structure functions that enter the triple differential distribution
for $b\to c\ell\bar \nu$ have become available only recently
\cite{trott,Uraltsev:2004in}.  In the limit of massless leptons, just
three of these structure functions are actually relevant, but their
knowledge is necessary in order to compute the moments of a generic
distribution with arbitrary experimental cuts and it is therefore
phenomenologically important.  At the level of physical observables,
but not of structure functions, semileptonic $b\to u$ decays can be
obtained as the $m_c\to 0$ limit of our results.  In this case the expressions
simplify considerably, and the complete structure functions are known
since a few years \cite{defazio}.

Beyond the one-loop approximation, several results are already
available.  In particular, the $O(\alpha_s^2 \beta_0)$ ({i.e.}\ the
so-called BLM~\cite{blm}) corrections to the total width have been
computed in \cite{Luke:1994yc,Ball:1995wa}.  The equivalent
corrections to the lepton energy spectrum can be found in
\cite{gremmst}, while the first results of a comprehensive analysis of
hadronic moments have been reported in \cite{Uraltsev:2004in}.  The
BLM corrections are related to the running of the coupling constant in
the loop contributions, and are practically most easily computed using
one-loop diagrams calculated with a finite gluon mass
\cite{Ball:1995ni}.  Starting from this result, it is moreover
possible to compute and even resum higher order BLM contributions,
$O(\alpha_s^{n+1}\beta_0^n)$ \cite{Ball:1995wa,kolyablm}.  Complete
two-loop results are known for specific kinematic points only (at
zero, maximal, and intermediate recoil \cite{czarnecki}), from which
one can interpolate over the entire $q^2$ range and estimate the
correction to the total width, which turns out to be well approximated
by the BLM result.  Recently, even three-loop $O(\alpha_s^3)$
corrections have become available for a particular kinematic
configuration, the {\it extreme zero-recoil} limit
\cite{Archambault:2004zs}.

In this paper, we calculate analytically the complete one-loop
corrections to the five structure functions relevant for the decay
$b\to c \ell \bar \nu$, paying particular attention to the proper
subtraction of the soft singularities, which is important for an
accurate numerical evaluation.  In this respect, our calculation
differs from \cite{trott,Uraltsev:2004in}.  In fact, our aim is to
provide a reliable tool to compute the QCD corrections to the moments
of $B\to X_c \ell \bar\nu$ in the most general experimental setting,
including the case $\ell=\tau$.  Whenever possible, we give our
results in analytic form; this in particular applies to the five
structure functions and to a few moments with an arbitrary cut on the
lepton energy. Our calculation represents an important check of
previous results and provides new analytic formulae that allow for an
easy and accurate numerical implementation. We find good agreement
with previous results whenever a comparison is possible.

Using the technique of refs. \cite{Ball:1995ni,blmvcb}, we also compute the
$O(\alpha_s^n \beta_0^{n-1})$ corrections to the triple differential
distribution and to its moments.  To this end, we have performed
analytically the calculation of the structure functions with a finite
gluon mass, and integrate numerically the results over the gluon mass,
applying the technique to the differential rate.  We present numerical
results for the one-loop and the BLM corrections in a few cases of
practical interest.

This paper is organized as follows: in Section~\ref{gamma} we set our
notation, briefly describe the calculation of the next-to-leading
order contributions, and discuss the subtraction of the soft
singularity; we also report at the end of this section the complete
expressions for the one-loop corrections to the structure functions.
In Section~\ref{BLMsect} we describe the technique used to calculate
the BLM corrections.  Section~\ref{numer} contains some reference
numerical result and a comparison with the literature.  Finally, we
briefly summarize our results in Section~\ref{summary}.  Details on
the kinematics involved to next-to-leading order are given in
Appendix~\ref{kinematics}, while in Appendix~\ref{virtual} we
illustrate the computation of the one-loop corrections.
Appendix~\ref{app_moments} gives analytic expressions for a few
moments of the hadronic invariant mass and energy with an arbitrary
cut on the lepton energy.

\section{The differential decay width to next-to-leading order}
\label{gamma}
The differential rate for the process
$$b(p)\to \ell(p_\ell)+\bar\nu(p_{\bar\nu})+X_c(r)$$
is given by
\beq
\frac{d\Gamma}{dq^2 du \,  dE_\ell}
=\frac{G_F^2\abs{V_{cb}}^2}{16\pi^4m_b^2}
\,L_{\mu\nu}(p_\ell,p_{\bar\nu})\,W^{\mu\nu}(p,q),
\eeq
where $q=p_\ell+p_{\bar\nu}$, $r=p-q$, $u=r^2-m_c^2$,
$E_\ell$ is the charged lepton energy
in the rest frame of the decaying $b$ quark, and for massless neutrinos
\beq
L^{\mu\nu}(p_\ell,p_{\bar\nu})
=p_\ell^\mu\, p_{\bar\nu}^\nu-p_\ell  p_{\bar\nu}\, g^{\mu\nu}
+p_\ell^\nu\, p_{\bar\nu}^\mu-i\epsilon^{\mu\alpha\nu\beta}{p_\ell}_\alpha
{p_{\bar\nu}}_\beta.
\eeq
Details on the kinematics involved are given in Appendix~\ref{kinematics}.
The decay rate is often expressed in terms of   
$q_0=(m_b^2+q^2-u-m_c^2)/(2m_b)$,
instead of $u$, which we have chosen for later convenience. 
The hadronic tensor $W^{\mu\nu}$ has the following, general decomposition,
based on Lorentz covariance and $CP$ symmetry:
\bea
W^{\mu\nu}(p,q)&=&-W_1(\hat q^2,\hat u)\,g^{\mu\nu}
+W_2(\hat q^2,\hat u)\,v^\mu v^\nu
+iW_3(\hat q^2,\hat u)\epsilon^{\mu\nu\rho\sigma} v_\rho \hat q_\sigma
\nonumber\\
&&+W_4(\hat q^2,\hat u)\,\hat q^\mu \hat q^\nu 
+W_5(\hat q^2,\hat u)\,(v^\mu \hat q^\nu+v^\nu \hat q^\mu),
\label{formfactors}
\eea
where $v=p/m_b$,  $\hat q=q/m_b$, and $\hat u=u/m_b$ (in the following, 
we denote with a hat 
dimensionless quantities, normalized to the $b$ quark mass).
In the limit of massless leptons the terms proportional to $W_4$ and 
$W_5$ do not contribute to the decay rate, since in that case $q_\mu
L^{\mu\nu}=q_\nu L^{\mu\nu}=0$. One finds
\beq
\frac{L_{\mu\nu}\,W^{\mu\nu}}{m_b^2}
=\hat q^2\,W_1(\hat q^2,\hat u)
-\left[2(v\hat p_\ell)^2-2v\hat p_\ell\,v\hat q
+\frac{\hat q^2}{2}\right]\,W_2(\hat q^2,\hat u)
+\hat q^2\left(2v\hat p_\ell-v\hat q\right)\,W_3(\hat q^2,\hat u).
\label{lw}
\eeq
The squared matrix element depends on
the charged lepton energy $E_\ell$ only through $vp_\ell=E_\ell$.

The leading contribution to the hadronic tensor $W^{\mu\nu}$
describes the process
\beq
b(p)\to \ell(p_\ell)+\bar\nu(p_{\bar\nu})+c(p')
\label{bc0}
\eeq
at the tree level. One finds
\beq
W_{(0)}^{\mu\nu}(p,q)=\left[-T_1\,g^{\mu\nu}
+T_2\,v^\mu v^\nu
+iT_3\,\epsilon^{\mu\nu\alpha\beta}v_\alpha\hat q_\beta
+T_4\,\hat q^\mu \hat q^\nu
+T_5(\hat q^\mu v^\nu+\hat q^\nu v^\mu)\right]
\,\pi\,\delta(\hat u),
\label{W0}
\eeq
where
\beq
T_1(\hat q^2)=-\hQt
\qquad
T_2(\hat q^2)=4
\qquad
T_3(\hat q^2)=2
\qquad
T_4(\hat q^2)=0
\qquad
T_5(\hat q^2)=-2,
\label{Ti}
\eeq
and we have defined
\beq
\hQt=\hat q^2-1-\hmct;\qquad
\hmct= \frac{m_c^2}{m_b^2}.
\eeq

The hadronic tensor receives contributions of order $\as$ from
one-gluon emission at the tree level,
\beq
b(p)\to \ell(p_\ell)+\bar\nu(p_{\bar\nu})+c(p')+g(k),
\label{bcg}
\eeq
integrated over the charm-gluon phase space (in this case, $r=p'+k$),
and from one-loop virtual corrections to the process in eq.~(\ref{bc0}):
\beq
W_{(1)}^{\mu\nu}(p,q)=W_{\rm (1)R}^{\mu\nu}(p,q)
+W_{\rm (1)V}^{\mu\nu}(p,q).
\eeq
The computation can be performed either by computing the relevant
squared matrix elements and integrating over the corresponding phase
space measures, or by taking the imaginary part of the $bW^*\to b W^*$
forward amplitude.  We calculated the structure functions in both ways; in
Appendix~\ref{virtual} we give some details about the first method only.
 We specifically consider here only the weak decay structure 
functions and assume the corresponding kinematics, {\it i.e.}\,  
$m_b^2\!\ge\! q^2\!\ge\! 0$  and $q_0\!\ge\! \sqrt{q^2}$. 

Let us begin with the contribution from real emission.
We have performed the computation with the help of the algebraic manipulation
programs Maxima~\cite{maxima} and Mathematica~\cite{math} for
an arbitrary gluon mass $\mg$. This is necessary for two reasons:
first, the gluon mass regulates the logarithmic 
infrared divergence that arises in the phase space integration from the soft
part of the gluon spectrum, and second, the full dependence of the one-loop
corrections on the gluon mass allows us to compute the 
order-$(\as^n\beta_0^{n-1})$ corrections.
The calculation is straightforward, and is already
discussed in the literature in different contexts~\cite{jezabek};
the only point that deserves
a careful discussion is the isolation of the infrared-singular term,
that is canceled by a similar contribution from virtual diagrams.
Both for conceptual reasons and for the numerical implementation,
it is important that the final result is manifestly
independent of the gluon mass in the $m_g\to 0$ limit.

While tree-level and virtual contributions are characterized by a 
$\delta(u)$ that defines their kinematic structure 
(production of an on-shell charm quark without gluons), the contributions from 
real emission diagrams are defined for values of 
the variable $\hat u$ between\footnote{On general grounds, 
the structure functions are actually proportional to 
$\theta(\hat u-\hat u_-)$. In the following,
we omit all $\theta$ functions from the explicit expressions, 
as it should not cause confusion.  }
\beq
\hat u_-=2\hmg\hmc+\hmg^2\qquad {\rm and}\qquad
\hat u_+=(1-\sqrt{\hat q^2})^2-\hmct.
\eeq
In this range, the real emission contributions to the structure
functions are regular as $\hmg\to 0$, but practical applications
require an integration over $\hat u$. In the physical limit $\hmg\to
0$, when the integration is performed close to the kinematic boundary
$\hat u=\hat u_-=0$ the real emission contributions develop a
$\log\hmg$ singularity. As mentioned above, this singularity is
canceled in physical quantities by a corresponding term originating
from the virtual corrections.  Since the singularity is localized at
$\hat u=0$, it is possible to extract it and write it in the form of a
term proportional to $\delta(\hat u)\,\log\hmg$, making the
cancellation of the infrared divergences explicit already {\it before}
integration.

This can be done as follows. The full result can be written as
\beq
W_{\rm (1)R}^{\mu\nu}(p,q)=W_{\rm (1)R,reg}^{\mu\nu}(p,q)
+W_{\rm (1)R,sing}^{\mu\nu}(p,q)
\eeq
where we have isolated in $W_{\rm (1)R,sing}^{\mu\nu}(p,q)$ all the
terms that give rise to a soft singularity when integrated over
$u$. There are obviously different ways to perform the splitting
between the two above contributions, that differ in $W_{\rm
(1)R,sing}^{\mu\nu}$ by terms that do not give rise to singularities
upon integration over $u$.  Let us first concentrate on
$W_{\rm (1)R,sing}^{\mu\nu}$, which we define as
\beq
W_{\rm(1)R,sing}^{\mu\nu}(p,q)=C_F\,\as\,\left[
\hQt\,g^{\mu\nu}+4\,v^\mu v^\nu
+2i\,\epsilon^{\mu\nu\alpha\beta}\,v_\alpha \hat q_\beta
-2(v^\mu \hat q^\nu+v^\nu \hat q^\mu)\right]\,D(\hat q^2,\hat u,\hmg^2),
\label{Wrealdiv}
\eeq
where $C_F=4/3$, and
\bea
\label{Dhatdef}
&&D(\hat q^2,\hat u,\hmg^2)=\frac{1}{\hat u}\frac{\hQt}{\sqrt{\lamb}}\,
\log\frac{\hat t_+}{\hat t_-}
-\frac{1}{\sqrt{\lamb}}\frac{\hat t_+-\hat t_-}{\hat t_+\hat t_-}
-\frac{\hmct}{\hat u^2}\,\frac{\hat t_+-\hat t_-}{\sqrt{\lamb}},
\\
&&\hat t_\pm=\hmg^2+\frac{(\hat u+\hmg^2)(\hQt-\hat u)
\pm\sqrt{\lamb}\sqrt{\lamc}}{2(\hat u+\hmct)}
\label{tbounds}
\\
&&\lamb=(\hQt-\hat u)^2-4(\hat u+\hmct)\\
&&\lamc=(\hat u-\hmg^2)^2-4\hmg^2\hmct.
\eea
In the limit $\hmg\to 0$ 
\bea
D(\hat q^2,\hat u,0)&=&\frac{D_0(\hat q^2,\hat u)}{\hat u}
\\
D_0(\hat q^2,\hat u)&=&
\frac{\hQt}{\sqrt{\lamb}}\,
\log\frac{\hQt-\hat u+\sqrt{\lamb}}{\hQt-\hat u-\sqrt{\lamb}}
-\frac{\hat u+2\hmct}{\hat u+\hmct},
\label{Dtilde}
\eea
so that the integral over $\hat u$ is logarithmically divergent.
The two above equations are  sufficient to
isolate the divergent term, but not to evaluate the finite
contribution.  The problem is  that $D(\hat q^2,\hat u,\hmg)$ 
is not analytic when  $\hat u=\hat u_-$ and $\hmg= 0$:
indeed, 
\bea
D(\hat q^2,\hat u,\hmg^2)&=& 
D_s(\hat q^2,\hat u,\hmg^2)+O(\hat u-\hat u_-,\hmg)
\\
D_s(\hat q^2,\hat u,\hmg^2)&=&\frac{1}{\hat u}\frac{\hQt}{\sqrt{\lamb^0}}\,
\log\frac{1+\frac{\sqrt{\lamb^0}}{\hQt}
\frac{\sqrt{\hat u^2-\hat u_-^2}}{\hat u}}{1-\frac{\sqrt{\lamb^0}}{\hQt}
\frac{\sqrt{\hat u^2-\hat u_-^2}}{\hat u}}
-\frac{\sqrt{\hat u^2-\hat u_-^2}}{\hat u^2+\frac{\lamb^0 \hat u_-^2}
{4\hmct}}
-\frac{\sqrt{\hat u^2-\hat u_-^2}}{\hat u^2},
\eea
where $\lamb^0=\hQt^2-4\hmct$. The  integration over $\hat u$
becomes non-trivial when it starts from the endpoint $\hat u=\hat u_-$.
We now add and subtract to $D(\hat q^2,\hat u,\hmg^2)$ a term proportional
to $\delta(\hat u)$:
\beq 
D(\hat q^2,\hat u,\hmg^2)=
\left[D(\hat q^2,\hat u,\hmg^2)-A(\hat q^2,\hmg)\,\delta(\hat u)
\right]+A(\hat q^2,\hmg)\,\delta(\hat u)
\label{addsub}
\eeq
where we have defined 
\beq
A(\hat q^2,\hmg)=\int_{\hat u_-}^{\hat u_+} d\hat u\,
D_s(\hat q^2,\hat u,\hmg^2).
\label{Adef}
\eeq
The $\hat u$ integration of the term in square brackets in eq.~(\ref{addsub})
is  regular for $\hat \mg\to 0$ 
over any range, while the  soft singularity is entirely contained in 
$A(\hat q^2,\hmg)$. For instance, 
for a generic test function $F(\hat u)$ regular
in $\hat u=0$, we have
\beq
\int_{\hat u_-}^{\hat u_+} d \hat u \, F(\hat u) D(\hat q^2,\hat u,\hmg^2)
=
\int_{\hat u_-}^{\hat u_+} d \hat u \left[F(\hat u) D(\hat q^2,\hat u,\hmg^2)-
F(0) D_s(\hat q^2,\hat u,\hmg^2)\right]+ 
F(0)\,  A(\hat q^2,\hmg),
\label{integr}
\eeq
The integration range in
eq.~(\ref{integr}) is irrelevant and may be restricted by the
shape of the test function. We have chosen it to coincide with the
allowed kinematic range $\hat u_-\leq\hat u\leq\hat u_+$ so that for
$F(\hat u)=1$ the integral in eq.~(\ref{integr}) is the one which is
relevant for the computation of the total rate.

One can make notation more compact by defining
\beq
D(\hat q^2,\hat u,\hmg^2)=\left[D(\hat q^2,\hat u,\hmg^2)\right]_+
+A(\hat q^2,\hmg)\,\delta(\hat u),
\label{DpA}
\eeq
where the plus prescription is\footnote{This is a generalization of the 
standard 'plus' distribution.}
\beq
\int_{\hat u_-}^{\hat u_+}d\hat u\,
\left[D(\hat q^2,\hat u,\hmg^2)\right]_+\,F(\hat u)
=\int_{\hat u_-}^{\hat u_+}d\hat u\,
\left[D(\hat q^2,\hat u,\hmg^2)F(\hat u)-D_s(\hat q^2,\hat u,\hmg^2)
F(0)\right].
\label{Ddistr}
\eeq
In the limit $\hmg\to 0$, we have
\beq
\left[D(\hat q^2,\hat u,\hmg^2)\right]_+=
D_0(\hat q^2,\hat u)\,\left(\frac{1}{\hat u}\right)_+ ,
\label{ddistr4}
\eeq
where the distribution $(1/\hat u)_+$ is defined by
\beq
\int_0^{\hat u_+}d\hat u\,\left(\frac{1}{\hat u}\right)_+ f(\hat u)=
\int_0^{\hat u_+}\frac{d\hat u}{\hat u}\,[f(\hat u)-f(0)].
\eeq 
As anticipated, the last term in eq.~(\ref{DpA}) 
has the same kinematic structure as the tree-level and virtual
contributions.
The integral $A(\hat q^2,\hmg)$ can be computed analytically for 
generic values of $\hmg$; in the small-$\hmg$ limit we have
\bea
A(\hat q^2,\hmg)&=&
\left(2-\frac{1}{a}\log\frac{1+a}{1-a}\right)
\log\frac{\hmg\hmc}{\hat u_+}
+\frac{1}{2a}\Bigg[
  {\rm Li}_2\left(\frac{2a}{a-1}\right)
 -{\rm Li}_2\left(\frac{2a}{a+1}\right)
\Bigg]
\nonumber\\
&&
+\frac{1}{2a}\log\frac{1+a}{1-a}+1+O(\hmg^2),
\label{amg0}
\eea
where  $a=\sqrt{\lamb^0}/\hQt$. Finally we have
\bea
 W^{\mu\nu}_{\rm (1)R,sing}(p,q)&=&C_F\,\as\,
\left[\hQt\,g^{\mu\nu}+4\,v^\mu v^\nu
+2i\,\epsilon^{\mu\nu\alpha\beta}\,v_\alpha\hat q_\beta
-2(v^\mu\hat q^\nu+v^\nu\hat q^\mu)
\right]\,\left[D(\hat q^2,\hat u,\hmg^2)\right]_+
\nonumber\\
&+&
\frac{C_F\,\as}{\pi}\,W_{(0)}^{\mu\nu}(p,q)\,A(\hat q^2,\hmg).
\label{Wrealsing}
\eea

We now consider the regular part of the real emission contribution. It can be 
written in terms of structure functions as in eq.~(\ref{formfactors}):
\bea
 W^{\mu\nu}_{\rm (1)R, reg}(p,q)&=&C_F\,\as\,\Big\{
-R_1(\hat q^2,\hat u)\,g^{\mu\nu}
+R_2(\hat q^2,\hat u)\,v^\mu v^\nu
+iR_3(\hat q^2,\hat u)\,
\epsilon^{\mu\nu\alpha\beta}\,v_\alpha\hat q_\beta
\nonumber\\
&&\phantom{aaaaa}+R_4(\hat q^2,\hat u)\,\hat q^\mu\hat q^\nu
+R_5(\hat q^2,\hat u)\,(v^\mu\hat q^\nu+\hat q^\mu v^\nu)
\Big\}.
\label{Wreg}
\eea
The analytic expressions of the structure functions $R_i$ for $\hmg=0$ are
\bea
R_1(\hat q^2,\hat u)&=&
\frac{\hat u\left( 3\hat u + 2\rho  \right) +
\omega\left( 3\hat u + 4\rho  \right) }
   {4{\left( \hat u + \rho  \right) }^2} + 
  \frac{\hat u^2( \omega - \hat u) }
   {2\left( \hat u + \rho  \right) \lamb }
+\frac{6\hat u}{\lamb }
\nonumber\\
&&+\frac{4\hat u^2-(6\hQt+\lamb)\hat u+2\lamb\hQt}{2\lamb \sqrt{\lamb}}
\,\log\tau 
\label{Rg}
\\
R_2(\hat q^2,\hat u)&=&
\frac{2\left[   \hat\omega\hat u (8-9\hat u)  -8{\hat\omega}^3 +
15{\hat\omega}^2\hat u  + 2\hat u (\hat u^2-8  \hat u+8)
\right] }{{\lamb}^2} +
   \frac{64\left( 1 + \hat\omega - \hat u \right) \rho }
   {{\lamb}^2} 
\non\\&&
+ \frac{{\hat\omega}\hat u  (\hat\omega-\hat u)^3 }{{\lamb}^2
     {\left( \hat u + \rho  \right) }^2} - 
  \frac{2\left( 2{\hat\omega}^4 + {\hat\omega}^2\hat u (7+4\hat u) - 
       5{\hat\omega}^3\hat u - 9\hat\omega\hat u^2
 + 2\hat u^3 - \hat\omega\hat u^3
       \right) }{{\lamb}^2\left( \hat u + \rho  \right) }
\label{Rpp}
\\&&+\frac{2}{\lamb^2 \sqrt{\lamb}} \Bigg[
 2{\lamb}^2 + 
      \lamb ( 26\hat u - 2\hat u^2 + 
         \hat\omega ( 8 + 3\hat u)  + 16\rho  ) 
\\
&&       - 6\hat u\left( 3\hat u^2 + 2\hat u ( \rho -5 )  - 
         12\rho  - \hat\omega ( 3 + 4\hat u + 3\rho)  \right)
\Bigg]
\,\log\tau 
\nonumber\\
R_3(\hat q^2,\hat u)&=&
-\frac{(\hQt-2)\hat u-\hQf-4(\hQt+\hmct)+\lamb}
{\lamb\sqrt{\lamb}}\,\log\tau 
-\frac{\rho }{2{\left( \hat u + \rho  \right) }^2} 
\nonumber\\
&&
- \frac{3\hat\omega + \rho }
   {2\left( \hat\omega + \rho  \right) 
     \left( \hat u + \rho  \right) } - 
  \frac{8\hat\omega + 5\hat\omega^2 - 
     3\hat\omega \hat u + 4\rho  + 4\hat\omega\rho  - 
     2\hat u \rho }{\lamb
     \left( \hat\omega + \rho  \right) }
\label{Rep}\\
R_4(\hat q^2,\hat u)&=&
-\frac{8\left( 2{\hat\omega}^2 - 2\hat u - 5\hat\omega \hat u + 
       3\hat u^2 \right) }{{\lamb}^2} + 
  \frac{64\rho }{{\lamb}^2} + 
  \frac{\hat u^2 ({\hat\omega}-\hat u)^3 }{{\lamb}^2
     {\left( \hat u + \rho  \right) }^2} 
-\left[\frac{3\hat u(2\hat u-3\hQt)}{\lamb}\right.\\
&&
+\hat u-2\hQt\Bigg]
\frac{4\log\tau}{\lamb\sqrt{\lamb}}
+ \frac{2\hat u\left( \hat u^3-3{\hat\omega}^2(1-\hat u) - {\hat\omega}^3 + 
       \hat\omega (\hat u-3\hat u^2)  + 2\hat u^2   \right) }{{\lamb}^2
     \left( \hat u + \rho  \right) }
\non
\label{Rqq}
\\
R_5(\hat q^2,\hat u)&=&
\frac{2\left( 4{\hat\omega}^2 + 2{\hat\omega}^3 - 
       12\hat u -2\hat\omega\hat u (3-2\hat u) - 5{\hat\omega}^2\hat u + 
       2\hat u^2  - \hat u^3 \right) }{{\lamb}^2} 
- \frac{8\left( 8 + 2\hat\omega - \hat u \right) \rho }{{\lamb}^2} 
\nonumber\\&&
- \frac{\hat u({\hat\omega}   -\hat u)^3 ({\hat\omega}   +\hat u) }{2
     {\lamb}^2{\left( \hat u + \rho  \right) }^2} + 
  \frac{2{\hat\omega}^4 + {\hat\omega}^2\hat u (10+\hat u) - 
     4{\hat\omega}^3\hat u - 2\hat\omega\hat u^2 (5-  \hat u) - \hat u^4}{
     {\lamb}^2\left( \hat u + \rho  \right) }
\\&&
-\Bigg[
\lamb(16\hat u\! -\! \hat u^2 + \hat \omega(12 \!+\! \hat u) + 8\rho) + 
  6\hat u(8\hat u\! -\! \hat u^2 + \hat \omega(6 \!+ \!\hat u) + 12\rho)
\Bigg]\frac{\log\tau}{\lamb^2\sqrt{\lamb}} \non
\label{Rpq}
\eea
where
\beq
 \tau =\frac{\hQt-\hat u+ \sqrt{\lamb}}{\hQt-\hat u- \sqrt{\lamb}}.
\label{tbounds0}
\eeq

We now turn to the virtual contribution to the differential
width. We calculate the one-loop corrections to the decay of an on-shell $b$ 
quark into an on-shell $c$ quark and a virtual $W$ boson, parameterizing the 
result in terms of pole masses.
Some detail of the calculation is given in Appendix~\ref{virtual}.
The result has the form
\bea
W_{\rm (1)V}^{\mu\nu}(p,q)&=&
-C_F\,\as\,\Big\{
-V_1(\hat q^2,\hmg)\,g^{\mu\nu}+V_2(\hat q^2,\hmg)\,v^\mu v^\nu
+iV_3(\hat q^2,\hmg)\,\epsilon^{\mu\nu\alpha\beta}\,v_\alpha\hat q_\beta
\\
&&\phantom{aaaaa}+V_4(\hat q^2,\hmg)\,\hat q^\mu \hat q^\nu
+V_5(\hat q^2,\hmg)\,(v^\mu\hat q^\nu+\hat q^\mu v^\nu)\Big\}\,
\delta(\hat u)
-\frac{C_F\,\as}{\pi}\,W_{(0)}^{\mu\nu}(p,q)\,V_0(\hat q^2,\hmg),
\nonumber
\eea
where
\bea
V_0(\hat q^2,\hmg)&=&\frac{1}{2}\,(1+\hat I_0-2\hat J-2\hat I_1)
\nonumber\\
V_1(\hat q^2,\hmg)&=&
-\hQt\hat K
-(\hQf-\hQt -2\hmct)\hat I_x
-(\hQf-\hQt \hmct-2 \hmct)\hat I_y
\nonumber\\
V_2(\hat q^2,\hmg)&=&
-4\left[(\hQt+\hmct+1)\hat I_{xy}-\hQt(\hat I_x+\hat I_y)\right]
\nonumber\\
V_3(\hat q^2,\hmg)&=&
2\hat K+2(\hQt-1)\hat I_x+2(\hQt-\hmct)\hat I_y
\nonumber\\
V_4(\hat q^2,\hmg)&=&4  (\hat I_x-\hat I_{xy})
\nonumber\\
V_5(\hat q^2,\hmg)&=&
2\left[(\hQt+2)\hat I_{xy}-(\hQt+1)\hat I_x-(\hQt-\hmct)\hat 
I_y\right],
\label{Vi}
\eea
and the integrals $\hat I_0$, $\hat J$, etc. are defined in
Appendix~\ref{virtual}, eqs.~(\ref{I0Jdef}) and (\ref{intdef}).
In the limit $\hmg\to 0$ we have
\beq
\hat J=-1-\log\frac{\hmg^2}{\hmc};\qquad
\hat K=\frac{1}{2};\qquad
\hat I_0=\frac{\sqrt{\lamb^0}}{2\hat q^2}
\log\frac{z_+}{z_-}
 -\frac{1-\hmct}{2\hat q^2}\log\hmct
\eeq
\beq
\hat I_1=
\frac{\hQt}{\sqrt{\lamb^0}}\Bigg[
\frac{1}{2}\log\frac{z_+}{z_-}\log\frac{\hmg^2}{\sqrt{\rho}}
+\log z_-\log\left(\sqrt{\frac{z_-}{z_+}}\frac{1-z_+}{1-z_-}\right)
+{\rm Li}_2\left(1-\frac{z_+}{z_-}\frac{1-z_-}{1-z_+}\right)
-{\rm Li}_2\left(1-\frac{1-z_-}{1-z_+}\right)\Bigg]
\label{I1}
\eeq
\beq
\hat I_x=\frac{1}{2\hat q^2}\left[\log\frac{1}{\hmct}
-\frac{\hQt+2\hmct}{\sqrt{\lamb^0}}
\log\frac{z_+}{z_-}\right];\qquad
\hat I_y=\frac{1}{2\hat q^2}\left[\log\hmct
-\frac{\hQt+2}{\sqrt{\lamb^0}}
\log\frac{z_+}{z_-}\right]
\eeq
\beq
\hat I_{xy}=\frac{(1-\hmct)^2-\hat q^2(1+\hmct)}
{4\hat q^4\sqrt{\lamb^0}}
\log\frac{z_+}{z_-}
+\frac{\hmct-1}{4\hat q^4}\log\hmct
-\frac{1}{2\hat q^2},
\eeq
where
\beq
z_\pm=-\frac{\hQt\pm\sqrt{\lamb^0}}{2}=-\frac{\hQt}{2}
 \left(1\pm a\right).
\eeq
The soft singularity arising from loop integration
resides in the function $V_0(\hat q^2,\hmg)$,
\beq 
V_0(\hat q^2,\hmg)=
\left(2-\frac{1}{a}\log\frac{1+a}{1-a}\right)\log\hmg+\ldots,
\eeq
where the dots stand for terms regular in the limit $\hmg\to 0$. It 
cancels exactly the singularity we have found in eq.~(\ref{amg0}).

We conclude this section by summarizing our results for the structure functions
 to order $\as$ for massless gluon. We have found:
\bea
W_i(\hat q^2,\hat u)&=&\pi\Bigg\{
T_i(\hat q^2)\,\delta(\hat u)
+\frac{\as C_F}{\pi}\left[T_i(\hat q^2)\,S(\hat q^2)-V_i(\hat q^2)
\right]\,\delta(\hat u)
\non\\
&&\phantom{aaaaaaaaaaa}
+\frac{\as C_F}{\pi}\left[
T_i(\hat q^2)\,
D_0(\hat q^2,\hat u)\left(\frac{1}{\hat u}\right)_+
+R_i(\hat q^2,\hat u)\right]
\Bigg\},
\eea
where we have defined
\beq
S(\hat q^2)=\lim_{\hmg\to 0}\left[A(\hat q^2,\hmg)-V_0(\hat q^2,\hmg)\right].
\eeq
The $T_i$ are given in eq.~(\ref{Ti}),
$A$ in eq.~(\ref{amg0}),
$V_i$ in eq.~(\ref{Vi}),
$R_i$ in eqs.~(\ref{Rg}-\ref{Rpq}),
and $D_0$ in eq.~(\ref{Dtilde}).

Ref.~\cite{trott} also reports the results of the calculation of the
structure functions $W_i$. However,
a direct comparison of our results for $W_i$ with ref.~\cite{trott}
is possible only before subtraction of the soft singularity.
We agree with all the real emission contributions of ref.~\cite{trott},
but the virtual contributions to $W_3,W_4,W_5$ (in the notation of 
ref.~\cite{trott}) are different from our results. The discrepancy, however,
is such that our $V_1,V_2,V_3$ are reproduced correctly. Therefore,
a difference should be observed when computing the decay rate into
massive leptons only.

\section{Generalized BLM -- \boldmath $O(\as^n \beta_0^{n-1})$ corrections}
\label{BLMsect}
The full perturbative expansion
for the total rate can be written as
\beq
\Gamma=\Gamma_0\,
\left[1+a_1(0)\,\frac{\as}{\pi}+\left(\frac{\as}{\pi}\right)^2
\sum_{n=0}^\infty
\left(\frac{\as}{\pi}\right)^n\,a_{n+2}
\right],
\eeq
where 
\beq
\Gamma_0= \frac{G_F^2 m_b^5}{192\pi^3}
 (1-8\rho+8\rho^3-\rho^4-12\rho^2 \ln \rho)\label{Gamma0}
\eeq
is the tree-level semileptonic 
width, and the argument in the order-$\as$ correction $a_1$ is the gluon mass,
which is zero in the physical limit.
The coefficients $a_{n+2}$ are degree-$(n+1)$
polynomials in the number of light fermions $n_f$, or equivalently in
the first coefficient of the $\beta$ function for $\as$, $\beta_0=11-2 n_f/3$:
\beq
a_{n+2}=a_{n+2}^{\rm BLM}\,\beta_0^{n+1}+O(\beta_0^n).
\eeq
The dominant term $a_{n+2}^{\rm BLM}$ in the large-$\beta_0$ limit,
usually called the (extended) BLM approximation~\cite{blm},
can be computed in terms of the first perturbative coefficient
$a_1(\mg^2)$~\cite{Ball:1995ni}; in the $\overline{\rm MS}$ scheme 
for the coupling constant the result reads \cite{imprecated,Ball:1995ni}
\bea
&&a^{\rm BLM}_{n+2}=-\frac{1}{4^{n+1}}\sum_{k=0}^{n/2}
(-\pi^2)^k\left(\begin{array}{c}n+1\\2k+1\end{array}\right)
A_{nk}
\label{ApertBLM}
\\
&&A_{nk}=\int_0^\infty\frac{d\mg^2}{\mg^2}
\,\log^{n-2k}\frac{\mu^2}{\mg^2}
\left[a_1(\mg^2)-a_1(0)\frac{\mu^2}{\mu^2+\mg^2}\right];
\qquad \mu^2=m_b^2\,e^{5/3}.
\eea
We will now obtain an expression similar to eq.~(\ref{ApertBLM})
for the BLM correction to the differential rate.
As we have seen in the previous section, the
order-$\as$ correction to the total rate
is obtained by summing the virtual-soft contribution  
proportional to $\delta(u)$ and the real contribution:
\beq
a_1(\mg^2)=a_1^{\rm VS}(\mg^2)+a_1^{\rm R}(\mg^2).
\eeq
After integration over the lepton energy $E_\ell$, 
with or without kinematic cuts, we have
\bea
&&a_1^{\rm VS}(\mg^2)=\int_0^{\delta^2}dq^2\,
\frac{da_1^{\rm VS}(q^2,\mg^2)}{dq^2}
\\
&&a_1^{\rm R}(\mg^2)=\theta(\delta-\mg)\,
\int_0^{(\delta-\mg)^2}dq^2\,\int_{u_-}^{u_+}du\,
\frac{da_1^{\rm R}(q^2,u,\mg^2)}{dq^2du}
\eea
where we have defined $\delta=m_b-m_c$. The integration bounds on
$u=r^2-m_c^2$ are
\beq
u_-=(m_c+\mg)^2-m_c^2;\qquad u_+=(m_b-\sqrt{q^2})^2-m_c^2.
\eeq
The $u$ integration in the virtual-soft term is trivial, since it
is  proportional to $\delta(u)$. Also note that
the real emission term vanishes when $\mg>\delta$; this
is implicitly taken into account in the definition of $a_1^{\rm VS}$,
which contains the soft contribution of the real emission
term. We have
\bea
A_{nk}&=&\int_0^\infty\frac{d\mg^2}{\mg^2}
\,\log^{n-2k}\frac{\mu^2}{\mg^2}
\left[a_1^{\rm VS}(\mg^2)+a_1^{\rm R}(\mg^2)
-\frac{\mu^2}{\mg^2+\mu^2}\left(a_1^{\rm VS}(0)
+a_1^{\rm R}(0)\right)\right]
\nonumber\\
&=&\int_0^\infty\frac{d\mg^2}{\mg^2}
\,\log^{n-2k}\frac{\mu^2}{\mg^2}
\int_0^{\delta^2}dq^2\,
\left[\frac{da_1^{\rm VS}(q^2,\mg^2)}{dq^2}
-\frac{da_1^{\rm VS}(q^2,0)}{dq^2}\,
\frac{\mu^2}{\mg^2+\mu^2}\right]
\nonumber\\
&&+\Bigg[
\int_0^{\delta^2}\frac{d\mg^2}{\mg^2}
\,\log^{n-2k}\frac{\mu^2}{\mg^2}
\int_0^{(\delta-\mg)^2}dq^2\,\int_{u_-}^{u_+}du\,
\frac{da_1^{\rm R}(q^2,u,\mg^2)}{dq^2du}
\nonumber\\
&&
-\int_0^\infty\frac{d\mg^2}{\mg^2}
\,\log^{n-2k}\frac{\mu^2}{\mg^2}
\int_0^{\delta^2}dq^2\,\int_0^{u_+}du\,
\frac{da_1^{\rm R}(q^2,u,0)}{dq^2du}\,
\frac{\mu^2}{\mg^2+\mu^2}\Bigg].
\label{d1bbb2}
\eea
Now we observe that
\beq
\int_0^{\delta^2}d\mg^2
\int_0^{(\delta-\mg)^2}dq^2\,\int_{u_-}^{u_+}du
=\int_0^{\delta^2}dq^2\,\int_0^{u_+}du\int_0^{\Lambda^2}d\mg^2,
\eeq
with
\beq
\Lambda^2=\left(\sqrt{r^2}-m_c\right)^2.
\eeq
Thus,
\bea
A_{nk}&=&
\int_0^\infty\frac{d\mg^2}{\mg^2}\int_0^{\delta^2}dq^2\,
\left[\frac{da_1^{\rm VS}(q^2,\mg^2)}{dq^2}
-\frac{da_1^{\rm VS}(q^2,0)}{dq^2}\,
\frac{\mu^2}{\mg^2+\mu^2}\right]
\,\log^{n-2k}\frac{\mu^2}{\mg^2}
\nonumber\\
&&+
\int_0^{\delta^2}dq^2\,\int_0^{u_+}du
\int_0^{\Lambda^2}\frac{d\mg^2}{\mg^2}
\left[\frac{da_1^{\rm R}(q^2,u,\mg^2)}{dq^2du}
-\frac{da_1^{\rm R}(q^2,u,0)}{dq^2du}\,
\frac{\mu^2}{\mg^2+\mu^2}\right]
\,\log^{n-2k}\frac{\mu^2}{\mg^2}
\nonumber\\
&&
-\int_0^{\delta^2}dq^2\,\int_0^{u_+}du\,
\frac{da_1^{\rm R}(q^2,u,0)}{dq^2du}\,
\int_{\Lambda^2}^\infty\frac{d\mg^2}{\mg^2}\frac{\mu^2}{\mg^2+\mu^2}
\,\log^{n-2k}\frac{\mu^2}{\mg^2}.
\label{BLMint}
\eea
The $\mg^2$ integral in the last term can be
expressed in terms of polylogarithms; for example,
\bea
\int_{\Lambda^2}^\infty\frac{d\mg^2}{\mg^2}\frac{\mu^2}{\mg^2+\mu^2}
&=&\log\frac{\Lambda^2+\mu^2}{\Lambda^2}
\nonumber\\
\int_{\Lambda^2}^\infty\frac{d\mg^2}{\mg^2}\frac{\mu^2}{\mg^2+\mu^2}
\log\frac{\mu^2}{\mg^2}
&=&{\rm Li}_2\left(-\frac{\mu^2}{\Lambda^2}\right)
-\log\frac{\Lambda^2}{\mu^2}\log\frac{\mu^2+\Lambda^2}{\Lambda^2}.
\eea
From eq.~(\ref{BLMint}), we can read off an expression
for the BLM corrections to the differential rate:
\bea
\frac{dA_{nk}}{dq^2 du}&=&
\label{BLMdiff}
\int_0^\infty\frac{d\mg^2}{\mg^2}\delta(u)\,
\left[\frac{da_1^{\rm VS}(q^2,\mg^2)}{dq^2}
-\frac{da_1^{\rm VS}(q^2,0)}{dq^2}\,
\frac{\mu^2}{\mg^2+\mu^2}\right]
\,\log^{n-2k}\frac{\mu^2}{\mg^2}
\nonumber\\
&&+
\int_0^{\Lambda^2}\frac{d\mg^2}{\mg^2}
\left[\frac{da_1^{\rm R}(q^2,u,\mg^2)}{dq^2du}
-\frac{da_1^{\rm R}(q^2,u,0)}{dq^2du}\,
\frac{\mu^2}{\mg^2+\mu^2}\right]
\,\log^{n-2k}\frac{\mu^2}{\mg^2}
\nonumber\\
&&
-\frac{da_1^{\rm R}(q^2,u,0)}{dq^2du}\,
\int_{\Lambda^2}^\infty\frac{d\mg^2}{\mg^2}\frac{\mu^2}{\mg^2+\mu^2}
\,\log^{n-2k}\frac{\mu^2}{\mg^2}.
\eea
Eq.~(\ref{BLMdiff}) can be used for a numerical computation of the
BLM coefficients,
since each integral is separately convergent. Any kind of kinematic
cuts can be implemented using appropriate step-functions in the integrands.

All the one-loop results of Section~\ref{gamma}, as well as the 
above expressions, have been derived in the on-shell scheme for the 
$b$ and $c$ quark masses. The use of pole masses is known to induce large  
higher order perturbative corrections and is not adequate to precision studies.
In most practical $B$ physics applications (see {\it e.g.}\ \cite{moments})
it is therefore appropriate to employ well-defined 
short distance mass parameters, such as the so-called {\it kinetic} masses,
namely heavy quark parameters which are renormalized {\it a la}
Wilson and depend explicitly on a `hard' normalization scale $\mu$
which is chosen close  to 1 GeV \cite{kolyablm,kinetic}. 
The change of scheme required to express $O(\alpha_s^{n+1} \beta_0^n)$  
corrections in terms of short-distance parameters is elementary, once 
the relation between the latter and their on-shell counterpart is known 
at the very same order.

\section{Numerical results}
\label{numer}
The primary field of application of the results described in the
previous sections is the calculation of the perturbative
contributions to the first few moments of various differential decay
distributions.  The moments most frequently employed 
are those of the charged lepton
energy spectrum, and of the invariant mass of the hadronic system
in the final state~\cite{moments_exp}.
As experimental cuts are usually applied, the moments of phenomenological
interest are actually truncated. A typical case is 
a lower cut on the charged lepton energy. At the
one-loop level, a fully analytic approach is possible even with a cut on the
lepton energy, but
the integration over the phase space is a tedious task.
We have performed
the phase space integration analytically for several of the partonic
moments that represent the building blocks of the first three
hadronic mass moments
(see \cite{fls95,moments}), in the case of an arbitrary cut on the lepton
energy. A few representative examples of the results are reproduced in
Appendix~\ref{app_moments}. However, in general, a strategy for an
accurate and fast
numerical evaluation is necessary, especially when computing
higher order corrections in the BLM approximation.

We have implemented these calculations in a {\sc FORTRAN} code, which is
available from the authors upon request. The subtraction of soft
singularities described in Section~\ref{gamma} allows us to compute a generic
moment and to apply arbitrary kinematic cuts.

In this section we provide a few tables  of reference numbers obtained
with our code for massless leptons; they have been checked in a number of ways,
both analytical and numerical. All results are in the pole mass scheme.

\begin{table}[t]
\begin{center}
\begin{tabular}{|c|l|l|l|l|}
\hline
n & $L^{(0)}_n$ & $L^{(1)}_n$ & $L^{(2)}_n$ & $L^{(3)}_n$ \\
\hline
\hline
 0 &   1. &  $-$1.777588 &  $-$1.917079 &  $-$2.827995\\
 \hline
 1 &   0.307202 &  $-$0.551243 &  $-$0.617666 &  $-$0.938365\\
 \hline
 2 &   0.103000 &  $-$0.187687 &  $-$0.217405 &  $-$0.338429\\
 \hline
 3 &   0.036524 &  $-$0.067828 &  $-$0.080868 &  $-$0.128695\\
 \hline
\end{tabular}
\end{center}
\caption{\label{tab:l0} Leptonic moments without $E_\ell$ cut.}
\end{table}

In particular, we focus here on the leptonic moments,
\bea
L_n&=& \frac1{\Gamma_0}\int d q^2 d u\, d E_\ell 
\ \hat E_\ell^n  \ \frac{d^3 \Gamma}{d q^2 \,d u \, d E_\ell}
\\ \non &=&
L_n^{(0)}+\frac{ \alpha_s}{\pi} L_n^{(1)}+\frac{ \alpha^2_s\beta_0}
{\pi^2}L_n^{(2)}+ \frac{ \alpha^3_s\beta_0^2}
{\pi^3}L_n^{(3)}+...
\eea
and on the hadronic moments of the form
\bea
H_{ij}&=& \frac1{\Gamma_0}\int  d q^2 d u \,d E_\ell 
\ \hat u^i \ \label{had_mom}
\hat E_0^j \ \frac{d^3 \Gamma}{d q^2 \,d u \,d E_\ell} \\
\nonumber&=&
H_{ij}^{(0)}+\frac{\alpha_s}{\pi} H_{ij}^{(1)}+\frac{ \alpha^2_s\beta_0}
{\pi^2}H_{ij}^{(2)}+ \frac{ \alpha^3_s\beta_0^2}
{\pi^3}H_{ij}^{(3)}+...
\eea
where $\Gamma_0$ is the {\it total} tree-level width of eq.~(\ref{Gamma0}),
and  $E_0=(m_b^2+r^2-q^2)/2 m_b$ is the energy 
of the hadronic system at the partonic level in the $b$ quark rest frame.
We employ the on-shell masses $m_b=4.6$~GeV and  $m_c=1.15$~GeV. These 
numerical values do  not correspond  to the actual $b$ and $c$ pole masses, 
but they simplify the comparison with \cite{Uraltsev:2004in,moments}.  

In table~\ref{tab:l0} we present the results of $L_n^{(j)}$ for
$n=0,1,2,3$ and $j=0,1,2,3$ when no lower cut is applied on the
charged lepton energy.  The first row refers to the total width: we
reproduce the results of \cite{Nir:1989rm,Luke:1994yc,Ball:1995wa} for
the four entries.  The second to fourth rows refer to the first
few moments of the leptonic energy; we have checked that we reproduce with 
good accuracy the
results of~\cite{voloshin,gremmst} for the first two entries, while
the third and fourth entries have not been reported so far.

\begin{table}
\begin{center}
\begin{tabular}{|c|l|l|l|l|}
\hline
n & $L^{(0)}_n$ & $L^{(1)}_n$ & $L^{(2)}_n$ & $L^{(3)}_n$ \\
\hline
\hline
 0 &   0.814810 &  $-$1.439726 &  $-$1.600023 &  $-$2.407834\\
 \hline
 1 &   0.277642 &  $-$0.497511 &  $-$0.567236 &  $-$0.870422\\
 \hline
 2 &   0.097933 &  $-$0.178501 &  $-$0.208715 &  $-$0.326798\\
 \hline
 3 &   0.035615 &  $-$0.066185 &  $-$0.079299 &  $-$0.126665\\
 \hline
\end{tabular}
\end{center}
\caption{\label{tab:l1} Leptonic moments with $E_\ell>1$~GeV.}
\end{table}

Table~\ref{tab:l1} contains the same leptonic moments as
table~\ref{tab:l0}, but a 1~GeV lower cut 
on the lepton energy is applied. Here  the second entry in each row
confirms results by  \cite{gremmst}, while the third and fourth entries 
are new.

Reference values for the hadronic moments are given in
table~\ref{tab:h0} for the case without lepton energy cut, and in
table~\ref{tab:h1} for $E_\ell\ge 1$~GeV.  The values of
tables~\ref{tab:h0} and \ref{tab:h1} reproduce the tree-level and 
one-loop results
presented in refs.~\cite{fls95} and \cite{flcut}, taking into account
the different input values. The same holds for a few entries of the third 
column (second-order BLM) given in \cite{flcut}.
We always find acceptable numerical agreement with \cite{Uraltsev:2004in}.
More difficult is a comparison with the one-loop 
numerical results presented in 
\cite{trott}, as they refer to the physical hadronic invariant mass moments,
and are expressed in the 1S $b$ mass scheme. 
\begin{table}
\begin{center}
\begin{tabular}{|c|c|l|l|l|l|}
\hline
i & j & $H^{(0)}_{ij}$ &$H^{(1)}_{ij}$ &$H^{(2)}_{ij}$ &$H^{(3)}_{ij}$ \\
\hline
\hline
0 &  0 &   1. &  $-$1.777615 &  $-$1.917079 &  $-$2.827995\\
 \hline
 0 &  1 &   0.422009 &  $-$0.719047 &  $-$0.732727 &  $-$1.028913\\
 \hline
 0 &  2 &   0.183191 &  $-$0.291969 &  $-$0.275917 &  $-$0.362931\\
 \hline
 0 &  3 &   0.081475 &  $-$0.117703 &  $-$0.099753 &  $-$0.118659\\ \hline
 1 &  0 &   0. &   0.090094 &   0.161513 &   0.304195\\
 \hline
 1 &  1 &   0. &   0.047004 &   0.081646 &   0.148907\\
 \hline
 1 &  2 &   0. &   0.0250925 &   0.0421838 &   0.0743690\\
 \hline
 2 &  0 &   0. &   0.0091060 &   0.0124670 &   0.0163852\\
 \hline
 2 &  1 &   0. &   0.0053384 &   0.0071132 &   0.0090464\\
 \hline
 3 &  0 &   0. &   0.0018101 &   0.0021753 &   0.0023946\\
 \hline
\end{tabular}
\end{center}
\caption{\label{tab:h0} Hadronic moments without $E_\ell$ cuts.}
\end{table}

\begin{table}
\begin{center}
\begin{tabular}{|c|c|l|l|l|l|}
\hline
i & j & $H^{(0)}_{ij}$ &$H^{(1)}_{ij}$ &$H^{(2)}_{ij}$ &$H^{(3)}_{ij}$ \\
\hline
\hline
 0 &  0 &   0.814811 &  $-$1.439726 &  $-$1.600023 &  $-$2.407833\\
 \hline
 0 &  1 &   0.334079 &  $-$0.577255 &  $-$0.613259 &  $-$0.883227\\
 \hline
 0 &  2 &   0.141115 &  $-$0.234529 &  $-$0.235029 &  $-$0.320474\\
 \hline
 0 &  3 &   0.061196 &  $-$0.095945 &  $-$0.089096 &  $-$0.112775\\
 \hline
 1 &  0 &   0. &   0.057255 &   0.108008 &   0.212635\\
 \hline
 1 &  1 &   0. &   0.028504 &   0.052497 &   0.100739\\
 \hline
 1 &  2 &   0. &   0.0144748 &   0.0260422 &   0.0486621\\
 \hline
 2 &  0 &   0. &   0.0044430 &   0.0066138 &   0.0093668\\
 \hline
 2 &  1 &   0. &   0.0024224 &   0.0035595 &   0.0048962\\
 \hline
 3 &  0 &   0. &   0.0006414 &   0.0008931 &   0.0010505\\
\hline
\end{tabular}
\end{center}
\caption{\label{tab:h1} Hadronic moments with $E_\ell>1$~GeV.}
\end{table}

\section{Summary}
\label{summary}

We have performed a calculation of the $O(\alpha_s)$ and
$O(\alpha_s^{n}\beta_0^{n-1})$ corrections to the differential rate
for inclusive semileptonic $b$ decays.  We have calculated all five
structure functions; only three of them contribute to the decays into
massless leptons, while they are all relevant for $B\to X_c\tau \bar
\nu$.  Our one-loop calculation checks several existing results, while
the BLM calculation is far more general than those present in the
literature.  Since we explicitly subtract the soft singularity from
the gluon emission contribution, our results allow for an accurate
numerical implementation, which we provide in a {\sc FORTRAN} code,
available upon request.  The code computes $O(\alpha_s)$ and
$O(\alpha_s^{n}\beta_0^{n-1})$ corrections to the triple differential
distribution and integrates over the phase space with arbitrary
experimental cuts.  A few numerical results are presented in
Section~\ref{numer} for the case of massless leptons.

For what concerns the analyses of semileptonic moments,
relevant to the determination of $|V_{cb}|$, the present work permits
the evaluation of the perturbative contributions to hadronic moments at
the same level as for the leptonic moments, since all
$O(\alpha_s)$ and $O(\alpha_s^{n}\beta_0^{n-1})$ contributions can be
now computed. Practical applications have been touched upon in
\cite{Uraltsev:2004in} and will be expanded in a future publication.

\section*{Acknowledgements}

We are grateful to Giovanni Ossola and Oliver Buchmuller for helpful
discussions.  The work of P.~G.\ is supported in part by the EU grant
MERG-CT-2004-511156 and by MIUR under contract 2004021808-009, and
that of N.~U.\ by the NSF under grant PHY-0087419.

\appendix

\section{Kinematics}
\label{kinematics}
Let us consider the phase space element for the processes
\beq
b(p)\to \ell(p_\ell)+\bar\nu(p_{\bar\nu})+X_c,
\label{bclX}
\eeq
where the states $X_c$ contain a charm quark with momentum $p'$, plus
a collection of $n$ partons with momenta $k_1,\ldots,k_n$
and masses $m_1,\ldots,m_n$. It can be decomposed as
\beq
d\phi_{3+n}(p;p_\ell,p_{\bar\nu},p',k_i)=\frac{d q^2}{2\pi}\,
\frac{dr^2}{2\pi}\,d\phi_2(p;q,r)\,d\phi_2(q;p_\ell,p_{\bar\nu})
\,d\phi_{n+1}(r;p',k_i),
\label{dphi4}
\eeq
where
\bea
&&q^2_-\leq q^2 \leq q^2_+;\qquad
q^2_-=(m_\ell+m_\nu)^2=0;\quad q^2_+=(m_b-m_c-m_1-\ldots-m_n)^2
\\
&&r^2_-\leq r^2 \leq r^2_+;\qquad
r^2_-=(m_c+m_1+\ldots+m_n)^2;\quad r^2_+=(m_b-\sqrt{q^2})^2.
\eea
In the rest frame of the decaying $b$ quark,
the direction of the momentum $\vec q$ of the lepton pair
is irrelevant, and its modulus is fixed for fixed $q^2,r^2$.
Therefore, the first two-body phase space in eq.~(\ref{dphi4})
gives simply an overall factor:
\beq
d\phi_2(p;q,r)=\frac{d^3q}{(2\pi)^3 2q_0}\frac{d^3r}{(2\pi)^3 2r_0}
(2\pi)^4\delta^{(4)}(p-q-r)
=\frac{1}{4\pi}\frac{\abs{\vec q}}{m_b},
\label{dphi2i}
\eeq
where
\beq
\abs{\vec q}^2=\frac{\lambda(q^2,r^2,m_b^2)}{4m_b^2};
\qquad
\lambda(x,y,z)=x^2+y^2+z^2-2xy-2xz-2yz.
\eeq
We now turn to the phase space factor for the lepton pair. For a given
value of $q^2$, the kinematics is completely fixed by the energy of
one of the two leptons, for example the charged lepton energy
$E_\ell$:
\bea
d\phi_2(q;p_\ell,p_{\bar\nu})&=&\frac{d^3p_\ell}{(2\pi)^3 2E_\ell}
\frac{d^3p_{\bar\nu}}{(2\pi)^3 2E_{\bar\nu}}\,(2\pi)^4
\delta^{(4)}(q-p_\ell-p_{\bar\nu})
\nonumber\\
&=&\frac{1}{8\pi}
\frac{E_\ell\,dE_\ell\,d\cos\theta}
{|\vec q-\vec p_\ell|}\,
\delta(q^0-E_\ell-|\vec q-\vec p_\ell|).
\eea
where $\theta$ is the angle
formed by the directions of $\vec q$ and $\vec p_\ell$.
The delta function can be used to perform the angular integration.
We obtain
\beq
d\phi_2(q;p_\ell,p_{\bar\nu})=\frac{1}{8\pi}\frac{dE_\ell}{|\vec q|},
\label{dphi2lept}
\eeq
with
\beq
\cos\theta=\frac{2q^0E_\ell-q^2}{2|\vec q|E_\ell};
\qquad
q^0=\sqrt{q^2+\abs{\vec q}^2}=\frac{m_b^2+q^2-r^2}{2m_b}.
\eeq
From the condition $\abs{\cos\theta}\leq 1$ we get the limits on $E_\ell$:
\beq
E_\ell^-\leq E_\ell\leq E_\ell^+;
\quad
E_\ell^\pm=\frac{q^0\pm\abs{\vec q}}{2}.
\label{Elbounds}
\eeq
Collecting eqs.~(\ref{dphi4},\ref{dphi2i},\ref{dphi2lept})
we obtain
\beq
d\phi_{3+n}(p;p_\ell,p_{\bar\nu},p',k)=
\left[\frac{1}{128\pi^4m_b}\,dE_\ell\,dq^2\,dr^2\right]\,
d\phi_{n+1}(r;p',k_1,\ldots,k_n).
\eeq
For fixed $q^2$ and $E_\ell$, the variable $r^2$
is in one-to-one correspondence with the neutrino energy
$E_{\bar\nu}$:
\beq
r^2=(p-q)^2=m_b^2+q^2-2m_b(E_\ell+E_{\bar\nu}).
\eeq
Similarly, the invariant mass of the lepton pair
$q^2$ can be traded for the energy $E_{X_c}$ of the
hadronic system in the final state; they are related through
\beq
q^2=(p-r)^2=m_b^2+r^2-2m_bE_{X_c}.
\eeq

The differential decay width for the processes
in eq.~(\ref{bclX}) is given by
\beq
\frac{d^3\Gamma}{dq^2 dr^2 dE_\ell}
=\frac{1}{2}\frac{1}{2m_b}\frac{1}{128\pi^4m_b}\,
\sum_n\int d\phi_{n+1}(r;p',k_1,\ldots,k_n)\,
\abs{{\cal M}_n(p;p_\ell,p_{\bar\nu},p',k_i)}^2.
\eeq
At leading order in the weak interactions, we have
\beq
{\cal M}_n(p;p_\ell,p_{\bar\nu},p',k_i)=
-\frac{i}{q^2-m_W^2}
{\cal M}^\mu_n(p;q,p',k_i)
{\cal M}_\mu(q;p_\ell,p_{\bar\nu})
\simeq
\frac{i}{m_W^2}
{\cal M}^\mu_n(p;q,p',k_i)
{\cal M}_\mu(q;p_\ell,p_{\bar\nu}),
\eeq
where ${\cal M}^\mu_n(p;q,p',k_i)$ and
${\cal M}_\mu(q;p_\ell,p_{\bar\nu})$ are the amplitudes for the
processes
\bea
&&b(p)\to W^*(q)+X_c
\\
&&W^*(q)\to \ell(p_\ell)+\bar\nu(p_{\bar\nu})
\eea
respectively. We define a leptonic tensor $L^{\mu\nu}$ 
and a hadronic tensor $W^{\mu\nu}$ through
\bea
g^2\,L^{\mu\nu}(p_\ell,p_{\bar\nu})
&=&{\cal M}^\mu(q;p_\ell,p_{\bar\nu}){\cal M}^{*\nu}(q;p_\ell,p_{\bar\nu})
\label{Ldef}
\\
g^2\abs{V_{cb}}^2\,W^{\mu\nu}(p,q)
&=&\sum_n\int d\phi_{n+1}(r;p',k_1,\ldots,k_n)\,
{\cal M}_n^\mu(p;q,p',k_i){\cal M}^{*\nu}_n(p;q,p',k_i).
\label{Wdef}
\eea
In terms of $L^{\mu\nu}$ and $W^{\mu\nu}$, the decay rate is given by
\bea
\frac{d^3\Gamma}{dq^2 dr^2 dE_\ell}
&=&\frac{1}{2}\frac{1}{2m_b}\frac{1}{128\pi^4m_b}\,
\frac{g^4\abs{V_{cb}}^2}{m_W^4}
\,L_{\mu\nu}(p_\ell,p_{\bar\nu})\,W^{\mu\nu}(p,q)
\nonumber\\
&=&
\frac{G_F^2\abs{V_{cb}}^2}{16\pi^4m_b^2}
\,L_{\mu\nu}(p_\ell,p_{\bar\nu})\,W^{\mu\nu}(p,q).
\eea

Finally, we compute the phase space for the hadron system in the
cases $n=0,1$. For $n=0$, which is the case at order $\as^0$
and for the virtual contribution at order $\as$, we have simply
\beq
d\phi_1(r;p')=\frac{d^3p'}{(2\pi)^3 2p'_0}(2\pi)^4\delta^{(4)}(r-p')
=2\pi\,\delta(r^2-m_c^2).
\eeq
For $n=1$ we have a charm quark with momentum $p'$ and a gluon with
momentum $k$ in the final state.
Choosing the rest frame of the charm-gluon system, with
the 3-axis along the direction of $\vec p$, we find 
\beq
d\phi_2(r;p',k)=\frac{d^3p'}{(2\pi)^3 2p'_0}\frac{d^3k}{(2\pi)^3 2k_0}
(2\pi)^4\delta^{(4)}(r-p'-k)
=\frac{1}{8\pi}\frac{|\vec k|}{\sqrt{r^2}}\,d\cos\theta,
\eeq
with
\beq
|\vec k|^2=\frac{\lambda(r^2,m_c^2,\mg^2)}{4r^2};\qquad 
k_0=\sqrt{|\vec k|^2+\mg^2}=\frac{r^2-m_c^2+\mg^2}{2\sqrt{r^2}}.
\eeq
It will be convenient to introduce the variable
\beq
t=(p-k)^2-m_b^2=\mg^2-2(p_0k_0-|\vec p||\vec k|\cos\theta);
\qquad
d\cos\theta=\frac{dt}{2|\vec p||\vec k|},
\eeq
where
\beq
|\vec p|^2=\frac{\lambda(q^2,r^2,m_b^2)}{4r^2};\qquad
p^0=\sqrt{|\vec p|^2+m_b^2}=\frac{m_b^2+r^2-q^2}{2\sqrt{r^2}}
\eeq
because of the energy conservation constraint
\beq
\sqrt{|\vec p|^2+m_b^2}=\sqrt{r^2}+\sqrt{|\vec p|^2+q^2}.
\eeq
Hence
\bea
&&d\phi_2(r;p',k)=\frac{1}{8\pi}\frac{dt}{\sqrt{\lambda(q^2,r^2,m_b^2)}}
\label{dphi2c}
\\
&&t_+\leq t \leq t_-;\qquad
t_\pm=\mg^2-2p_0k_0\pm2|\vec p||\vec k|.
\eea

\section{Virtual corrections}
\label{virtual}
In this Appendix, we present in full detail the computation
of the order-$\as$ contribution to the differential rate due
to one-loop corrections to the process in eq.~(\ref{bc0}).
We have
\bea
\label{Wvirt}
W_{\rm (1)V}^{\mu\nu}(p,q)&=&
-i\,\frac{C_F\,g_s^2}{8}\,2\pi\delta(u)
\int\frac{d^d\ell}{(2\pi)^d}\frac{1}{\ell^2-\mg^2}
\frac{1}{\left[(\ell+p)^2-m_b^2\right]\left[(\ell+p')^2-m_c^2\right]}
\nonumber\\
&&\times{\rm Tr}
\Big[(\pps+m_c)\gamma_\rho(\ls+\pps+m_c)\gamma^\mu(1-\gamma_5)
(\ls+\ps+m_b)\gamma^\rho(\ps+m_b)\gamma^\nu(1-\gamma_5)
\nonumber\\
&&+(\ps+m_b)
\gamma_\rho(\ls+\ps+m_b)\gamma^\nu(1-\gamma_5)
(\ls+\pps+m_c)\gamma^\rho(\pps+m_c)\gamma^\mu(1-\gamma_5)\Big]
\nonumber\\
&&-i\,\left(Z_b+Z_c\right)\,W_{(0)}^{\mu\nu}(p,q),
\eea
where 
\beq
Z_i=\left.\frac{d\Sigma(p,m_i)}{d\ps}\right|_{\ps=m_i};
\qquad
\Sigma(p,m_i)=C_F\,g_s^2\,\int\frac{d^d\ell}{(2\pi)^d}
\frac{1}{\ell^2-\mg^2}\frac{\gamma_\rho(\ls+\ps+m_i)\gamma^\rho}
{(\ell+p)^2-m_i^2},
\eeq
and $p'=p-q$. With standard techniques, we obtain
\bea
\label{Wvirt2}
W_{\rm (1)V}^{\mu\nu}(p,q)&=&-iC_F\,g_s^2\,\int_0^1dx\int_0^{1-x}dy
\nonumber\\
&&\Bigg\{\frac{1}{2}\left[
\int\frac{d^d\ell}{(2\pi)^d}\frac{N_1^{\mu\nu}}{(\ell^2-M^2)^3}
+\int\frac{d^4\ell}{(2\pi)^4}\frac{N_2^{\mu\nu}}{(\ell^2-M^2)^3}
\right]\,\pi\delta(u)
\nonumber\\
&&+
\left[\int\frac{d^d\ell}{(2\pi)^d}
\frac{2-d}{(\ell^2-M_b^2)^2}
-\int\frac{d^4\ell}{(2\pi)^4}\frac{8m_b^2 x(1+x)}{(\ell^2-M_b^2)^3}\right]
\,W_{(0)}^{\mu\nu}(p,q)
\nonumber\\
&&+
\left[\int\frac{d^d\ell}{(2\pi)^d}
\frac{2-d}{(\ell^2-M_c^2)^2}
-\int\frac{d^4\ell}{(2\pi)^4}\frac{8m_c^2 x(1+x)}{(\ell^2-M_c^2)^3}\right]
\,W_{(0)}^{\mu\nu}(p,q)\Bigg\},
\nonumber
\eea
where we have collected in $N_1^{\mu\nu}$ all terms quadratic in the loop
momentum $\ell$, while $N_2^{\mu\nu}$ is $\ell$-independent. Furthermore,
\bea
M^2&=&m_b^2 x^2+m_c^2 y^2-\Qt  xy+\mg^2(1-x-y)
\\
M_i^2&=&m_i^2 x^2+\mg^2(1-x),
\eea
where $\Qt=q^2-m_b^2-m_c^2$.
Ultraviolet divergences are regulated by dimensional regularization,
with $d=4-2\epsilon$.
Using the properties of symmetric integration, we can replace
\beq
N_1^{\mu\nu}\to 8\,\ell^2\,\frac{(2-d)^2}{d}\,
\left[\Qt \, g^{\mu\nu}
+4\,p^\mu\, p^\nu
+2i\epsilon^{\mu\nu\alpha\beta}p_\alpha q_\beta
-2(q^\mu\, p^\nu+q^\nu\, p^\mu)
\right],
\eeq
and hence, by comparison with eq.~(\ref{W0}),
\beq
N_1^{\mu\nu}\,\pi\,\delta(u)
\to 8\,\ell^2\frac{(2-d)^2}{d}\,W_{(0)}^{\mu\nu}(p,q).
\eeq
Ultraviolet divergences are immediately seen to cancel, and setting $d=4$ 
we are left with
\beq
\label{Wrest}
W_{\rm (1)V}^{\mu\nu}(p,q)=-\frac{C_F\,\as}{2\pi}\Bigg\{
(1+I_0-2J)\,W_{(0)}^{\mu\nu}(p,q)
+\frac{\pi}{8}\,\delta(u)
\int_0^1dx\int_0^{1-x}dy\,
\frac{N_2^{\mu\nu}}{M^2}
\Bigg\},
\eeq
where we have defined
\bea
I_0&=&\int_0^1dx\int_0^{1-x}dy\,\log\frac{M^4}{M_b^2 M_c^2}
\nonumber\\
J&=&\int_0^1dx\,
x(1-x^2)
\left(\frac{m_b^2}{M_b^2}+\frac{m_c^2}{M_c^2}\right).
\label{I0Jdef}
\eea
After some algebra we find
\bea
N_2^{\mu\nu}&=&
-16 \Qt \left[\Qt \,g^{\mu\nu}+4\,p^\mu p^\nu
+2i\epsilon^{\mu\nu\alpha\beta}p_\alpha q_\beta
-2(p^\mu q^\nu+p^\nu q^\mu)\right]
\nonumber\\
&&
+16\left[\Qt M_0^2
+(\Qf -\Qt  m_b^2-2m_b^2 m_c^2)x
+(\Qf -\Qt  m_c^2-2m_b^2 m_c^2)y
\right]\,g^{\mu\nu}
\nonumber\\
&&
-64\left[(\Qt +m_c^2+m_b^2)xy-\Qt (x+y)\right]\,p^\mu p^\nu
\nonumber\\
&&
+32 \left[M_0^2+(\Qt -m_b^2)x+(\Qt -m_c^2)y\right]\,
i \epsilon^{\mu\nu\alpha\beta}p_\alpha q_\beta
\nonumber\\
&&
+64 m_b^2 x (1-y)\,q^\mu q^\nu
\nonumber\\
&&
+32\left[(\Qt +2 m_b^2)xy
-(\Qt +m_b^2)x-(\Qt -m_c^2)y\right]\,
(p^\mu q^\nu+p^\nu q^\mu),
\eea
where
\beq
M_0^2=m_b^2 x^2+m_c^2 y^2-\Qt xy.
\eeq
Our final result for the virtual contribution is therefore
\bea
W_{\rm (1)V}^{\mu\nu}(p,q)&=&
-C_F\,\as\,\Big\{V_0(\hat q^2,\hmg)\,
\left[\hQt\,g^{\mu\nu}
+4v^\mu v^\nu
+2i\epsilon^{\mu\nu\alpha\beta}v_\alpha\hat q_\beta
-2(\hat q^\mu v^\nu+\hat q^\nu v^\mu)\right]
\nonumber\\
&&\phantom{aaaaa}
-V_1(\hat q^2,\hmg)\,g^{\mu\nu}+V_2(\hat q^2,\hmg)\,v^\mu v^\nu
+iV_3(\hat q^2,\hmg)\,\epsilon^{\mu\nu\alpha\beta}\,v_\alpha\hat q_\beta
\nonumber\\
&&\phantom{aaaaa}+V_4(\hat q^2,\hmg)\,\hat q^\mu \hat q^\nu
+V_5(\hat q^2,\hmg)\,(v^\mu\hat q^\nu+\hat q^\mu v^\nu)\Big\}\,
\delta(\hat u),
\eea
where
\bea
V_0(\hat q^2,\hmg)&=&\frac{1}{2}\,(1+I_0-2J-2I_1)
\nonumber\\
V_1(\hat q^2,\hmg)&=&
-\frac{1}{m_b^2}\left[\Qt K
+(\Qf -\Qt  m_b^2-2m_b^2 m_c^2)I_x
+(\Qf -\Qt  m_c^2-2m_b^2 m_c^2)I_y
\right]
\nonumber\\
V_2(\hat q^2,\hmg)&=&
-4\left[(\Qt +m_c^2+m_b^2)I_{xy}-\Qt (I_x+I_y)\right]
\nonumber\\
V_3(\hat q^2,\hmg)&=&
2\left[K+(\Qt -m_b^2)I_x+(\Qt -m_c^2)I_y\right]
\nonumber\\
V_4(\hat q^2,\hmg)&=&4 m_b^2 (I_x-I_{xy})
\nonumber\\
V_5(\hat q^2,\hmg)&=&
2\left[(\Qt +2 m_b^2)I_{xy}-(\Qt +m_b^2)I_x-(\Qt -m_c^2)I_y\right]
\eea
and
\bea
K&=&\int_0^1dx\int_0^{1-x}dy\,\frac{M_0^2}{M^2}
\nonumber\\
I_1&=&\int_0^1dx\int_0^{1-x}dy\,\frac{\Qt }{M^2}
\nonumber\\
I_x&=&\int_0^1dx\int_0^{1-x}dy\,\frac{x}{M^2}
\nonumber\\
I_y&=&\int_0^1dx\int_0^{1-x}dy\,\frac{y}{M^2}
\nonumber\\
I_{xy}&=&\int_0^1dx\int_0^{1-x}dy\,\frac{xy}{M^2}
\label{intdef}
\eea
The integrals $I_1$ and $J$ are divergent for $\mg=0$; this is precisely
the soft singularity which is needed in order to cancel the
analogous divergent terms in the real emission contribution.
Also in this case, it is convenient to isolate the soft logarithm.
To this purpose, we perform the change in the integration variables
\bea
&&x=x_1 x_2
\\
&&y=x_1(1-x_2)\\
&&dx\, dy=x_1\,dx_1 dx_2;
\qquad
0\leq x_1\leq 1;\qquad 0\leq x_2\leq 1.
\eea
In terms of the new variables, we have
\beq
M^2=P(x_2)\,x_1^2+\mg^2(1-x_1),
\eeq
where
\beq
P(x_2)=m_b^2 x_2^2+m_c^2 (1-x_2)^2-\Qt x_2(1-x_2).
\eeq
Thus,
\bea
I_1&=&\Qt \int_0^1 dx_2\int_0^1dx_1\,\frac{x_1}{P(x_2)\,x_1^2+\mg^2(1-x_1)}
\nonumber\\
&=&\frac{\Qt }{2}\int_0^1 \frac{dx_2}{P(x_2)}
\int_0^1dx_1\,\frac{2P(x_2)x_1-\mg^2}{P(x_2)\,x_1^2+\mg^2(1-x_1)}
+\frac{\Qt }{2}\int_0^1 \frac{dx_2}{P(x_2)}
\int_0^1dx_1\,\frac{\mg^2}{P(x_2)\,x_1^2+\mg^2(1-x_1)}
\nonumber\\
&=&\frac{\Qt }{2}\int_0^1 \frac{dx_2}{P(x_2)}\log\frac{P(x_2)}{\mg^2}
+\frac{\Qt }{2}\int_0^1 \frac{dx_2}{P(x_2)}
\int_0^1dx_1\,\frac{\mg^2}{P(x_2)\,x_1^2+\mg^2(1-x_1)}.
\eea
The coefficient of $\log \mg^2$ can be computed explicitly. We find
\beq
-\frac{\Qt }{2}\int_0^1 \frac{dx_2}{P(x_2)}
=\frac{1}{2a}\log\frac{1+a}{1-a},
\eeq
where $a=\sqrt{\lamb^0}/\Qt $. Hence, 
\bea
I_1&=&\frac{1}{2a}\log\frac{1+a}{1-a}\,
\log\frac{\mg^2}{m_b m_c}
\nonumber\\
&&
+\frac{\Qt }{2}\int_0^1 \frac{dx_2}{P(x_2)}\log\frac{P(x_2)}{m_b m_c}
+\frac{\Qt }{2}\int_0^1 \frac{dx_2}{P(x_2)}
\int_0^1dx_1\,\frac{\mg^2}{P(x_2)\,x_1^2+\mg^2(1-x_1)}.
\eea
Similarly,
\bea
\int_0^1dx\,\frac{m_b^2x(1-x^2)}{m_b^2 x^2+\mg^2(1-x)}
&=&
\frac{1}{2}\int_0^1dx\,\frac{2m_b^2x-\mg^2}{m_b^2 x^2+\mg^2(1-x)}
+\frac{1}{2}\int_0^1dx\,\frac{\mg^2-2m_b^2 x^3}{m_b^2 x^2+\mg^2(1-x)}
\nonumber\\
&=&
\frac{1}{2}\log\frac{m_b^2}{\mg^2}
+\frac{1}{2}\int_0^1dx\,\frac{\mg^2-2m_b^2 x^3}{m_b^2 x^2+\mg^2(1-x)}
\eea
and therefore
\beq
J=-\log\frac{\mg^2}{m_b m_c}
+\frac{1}{2}\int_0^1dx\,
\left[\frac{\mg^2-2m_b^2 x^3}{m_b^2 x^2+\mg^2(1-x)}
+\frac{\mg^2-2m_c^2 x^3}{m_c^2 x^2+\mg^2(1-x)}\right].
\eeq

\section{Analytic expressions for one-loop moments}
\label{app_moments}
We present here a few examples of 
$O(\alpha_s)$ corrections to the moments of $u^i E_0^j$
at arbitrary values of the lower cut on the charged lepton energy.
Keeping in mind eqs.~(\ref{Gamma0}) and (\ref{had_mom}) we  define 
$$ 
M_{ij}=  \frac{\Gamma_0}{C_F \Gamma_u} H_{ij}^{(1)},
$$
with $\Gamma_u= G_F^2 |V_{cb}|^2 m_b^5/(192\pi^3)$, 
$\xi=2E_\ell^{\rm cut}/m_b$, and $L_\xi=\ln (1-\xi)$.

The moments with $i>0$ receive contributions from bremsstrahlung diagrams only
and are significantly easier to express in analytic form. Here are the lowest
moments:
\bea
M_{10}&=&\frac{-2873\,\rho^5}{1800} + \frac{\rho^4
     \left( 1 - 62\,\xi + 108\,{\xi}^2 - 93\,{\xi}^3 \right) }{24
     {\left( 1 - \xi \right) }^2} + 
  \frac{{\left( 1 - \xi \right) }^2\,
     \left( 91 - 298\,\xi - 52\,{\xi}^2 + 9\,{\xi}^3 \right) }{600}\non\\ 
&+& 
  \frac{\rho^3\,\left( 17 + 109\,\xi - 83\,{\xi}^2 + 18\,{\xi}^3 \right) }
   {6 - 6\,\xi} + \frac{\rho^2\,\left( -79 - 438\,\xi + 222\,{\xi}^2 + 
       25\,{\xi}^3 \right) }{18} 
\non\\&+& 
  \frac{\rho\left( 1 - \xi \right) 
     \left( 71 + 107\,\xi + 45{\xi}^2 - 51{\xi}^3 \right) }{24} + 
  \left[ 
\frac{\rho^3\,\left( 63 - 37\,\xi - 14\,{\xi}^2 \right) }
      {3 - 3\,\xi}  
\right.\non\\&+&
\frac{\rho^2
        \left( -146 + 195\,\xi - 60\,{\xi}^2 + 2\,{\xi}^3 \right) }{6} - 
     \frac{{\left( 1 - \xi \right) }^2\,
        \left( 96 + 17\,\xi - 62\,{\xi}^2 + 9\,{\xi}^3 \right) }{120}
\non\\&+& \left. 
     \frac{\rho^4\,\left( -60 + 137\,\xi - 118\,{\xi}^2 + 17\,{\xi}^3 
\right) }{24\,{\left( 1 - \xi \right) }^2} + 
     \frac{\rho\,\left( 9 - 20\,\xi + 18\,{\xi}^3 - 7\,{\xi}^4 \right) }{6} 
\right] L_\xi 
\non\\&+&
\left[ 4\rho^3\left(3-\xi \right)  - 
     \frac{\rho^4\,\xi}{2} + \rho^2\left( 8 - 9\,\xi + {\xi}^3 \right)  
\right] L_\xi \ln \frac{1-\xi}{\rho} +
\left[ \frac{7\,\rho^5}{15} - 
     \frac{\rho^3\left( 1 + 11\,\xi - 10\,{\xi}^2 \right) }{1 - \xi} +
\right.\non\\&+& 
 \left.
     \frac{\rho^4\left( 5 - 10\,\xi + 6\,{\xi}^2 \right) }
      {2\,{\left( 1 - \xi \right) }^2} + 
     \rho^2\left( \frac{5}{3} - 8\xi + 5\,{\xi}^2 - \frac{{\xi}^3}{6} 
\right)  + \frac{3\,\rho\,\left( 1 - 2\,{\xi}^3 + {\xi}^4 \right) }{2} \right]
 \ln \rho  \eea

\bea
M_{20}&=&\frac{6341\,\rho^6}{10800} + \frac{\rho^2( 1 \!- \!\xi ) 
     \left( -101 - 201\xi + 61{\xi}^2 + 41{\xi}^3 \right) }{48} + 
  \frac{\rho^3\left( 43 + 528\xi - 285{\xi}^2 + 47{\xi}^3 \right) }
   {54} \non\\
&+& \frac{{\left( -1 + \xi \right) }^3
     \left( -125 + 2013\xi - 474{\xi}^2 + 62{\xi}^3 \right) }{10800} + 
  \frac{\rho^4\left( 29 + 139\xi - 186{\xi}^2 + 134{\xi}^3 \right) }
   {-48 + 48\xi} 
\non\\&+&
 \frac{\rho{\left( 1 -\xi \right) }^2
     \left( -137 + 536\xi - 561{\xi}^2 + 262{\xi}^3 \right) }{600} + 
  \frac{\rho^5\left( 925 - 984\xi - 443{\xi}^2 + 734{\xi}^3 \right) }
   {600{\left( 1 - \xi \right) }^2}  
\non\\&+&
\left[ \frac{\rho^5\left( -2 + 3\xi \right) }{10} + 
     \frac{\rho^4\left( -6 + 3\xi + {\xi}^2 \right) }{2} - 
     \frac{2\rho^3\left( 2 - 3\xi + {\xi}^3 \right) }{3} \right] 
 L_\xi \ln \frac{1-\xi}{\rho}
\non\\&+&  
  \left( \frac{-7\rho^6}{45} + \frac{\rho^4
        \left( 7 - 19\xi + 12{\xi}^2 - 2{\xi}^3 \right) }{-4 + 
        4\xi} + \frac{\rho^5\left( -15 + 32\xi - 21{\xi}^2 + 
          3{\xi}^3 \right) }{10{\left( -1 + \xi \right) }^2} 
\right.\non\\&+& \left. 
     \frac{\rho^3\left( -38 + 24\xi - 24{\xi}^2 + 29{\xi}^3 \right) }
      {18} - \frac{5\rho^2\left( 1 - 2{\xi}^3 + {\xi}^4 \right) }{4} \
\right) \ln \rho  
\non\\&+&  
\left[ \frac{\rho^5
        \left( 866 - 1921\xi + 1484{\xi}^2 - 309{\xi}^3 \right) }
{600{\left( 1 - \xi \right) }^2} + 
     \frac{\rho^3\left( 176 - 207\xi + 84{\xi}^2 - 26{\xi}^3 \right) }
      {18} 
\right.\non\\&-& \left. 
 \frac{\rho{\left( 1 - \xi \right) }^2
        \left( -162 + 111\xi - 16{\xi}^2 + 7{\xi}^3 \right) }{120} + 
     \frac{\rho^2\left( -1 + \xi \right) 
        \left( 25 - 53\xi + {\xi}^2 + 9{\xi}^3 \right) }{12}  
\right.\non\\&-& \left. 
     \frac{{\left( 1 - \xi \right) }^3
        \left( 398 - 153\xi - 78{\xi}^2 + 13{\xi}^3 \right) }{1800} + 
     \frac{\rho^4\left( 84 - 63\xi - 22{\xi}^2 + 25{\xi}^3 \right) }
      {-24 + 24\xi} \right] L_\xi
\eea

\bea
M_{11}&=&\frac{-23\rho^6}{80} + \frac{\rho^3
     ( 6 + 20\xi - 16{\xi}^2 + 11{\xi}^3 ) }{6} + 
  \frac{\rho^5( 1457 - 4383\xi + 5481{\xi}^2 - 6356{\xi}^3 + 
       2604{\xi}^4 ) }{3600{\left( 1 - \xi \right) }^3} 
\non\\&+& 
  \frac{\rho( 1 \!- \!\xi )
     \left( 209 + 449\xi\! - 231{\xi}^2 \!- 76{\xi}^3 + 4{\xi}^4 
\right) }{240} + \frac{{( 1\! -\! \xi) }^2
     \left( 81 \!-\! 364\xi + 198{\xi}^2\! - \!48{\xi}^3 +\! 7{\xi}^4 
\right) }{1200} 
\non\\&+& 
 \frac{\rho^4\left( -49 - 340\xi + 570{\xi}^2 - 
       172{\xi}^3 + 37{\xi}^4 \right) }{48
     {\left( 1 - \xi \right) }^2} + 
  \frac{\rho^2\left( -149 + 468\xi - 702{\xi}^2 + 8{\xi}^3 + 
       87{\xi}^4 \right) }{144} 
\non\\&+&  
\left[ \frac{-\rho^5}{10} + 
     \rho^2{\left( -1 + \xi \right) }^3 + 
     \rho^3\left( -11 + 8\xi - {\xi}^2 \right)  + 
     \frac{\rho^4\left( -6 - 2\xi + {\xi}^2 \right) }{4} \right]
 L_\xi \ln \frac{1-\xi}{\rho}
 \non\\&+&   
  \left[ \frac{\rho^6}{10} + \frac{\rho^3
        \left( 2 + 44\xi - 10{\xi}^2 + 3{\xi}^3 \right) }{4} + 
     \frac{\rho^4\left( 1 + 4\xi - 6{\xi}^2 + {\xi}^3 - {\xi}^4 
\right) }{4{\left( 1 - \xi \right) }^2} 
\right.\non\\&+& \left. 
     \frac{\rho^2\left( 5 + 12\xi - 30{\xi}^2 + 7{\xi}^3 + 
          3{\xi}^4 \right) }{12} + 
     \frac{\rho^5\left( -23 + 57\xi - 39{\xi}^2 - 31{\xi}^3 + 
          24{\xi}^4 \right) }{120{\left( -1 + \xi \right) }^3}
\right.\non\\&-& \left.
     \frac{3\rho\left( -7 + 20{\xi}^3 - 15{\xi}^4 + 2{\xi}^5 
\right) }{40} \right] \ln \rho + 
  \left[ \frac{\rho^5\left( 1 - 63\xi + 213{\xi}^2 - 91{\xi}^3 
\right) }{300{\left( -1 + \xi \right) }^3} 
\right.\non\\&+& \left. 
     \frac{\rho^3\left( 40 - 184\xi + 38{\xi}^2 - 9{\xi}^3 \right) }
      {12} + \frac{\rho^2\left( 13 - 30\xi + 17{\xi}^2 + {\xi}^3 - 
          2{\xi}^4 \right) }{4} 
\right.\non\\&+& \left. 
     \frac{\rho( 1 \!- \!\xi ) 
        \left( 12 - 28\xi + 10{\xi}^2 + 4{\xi}^3 - {\xi}^4 \right) }
{12} - \frac{{( 1\! -\! \xi ) }^2
        \left( 526 - 218\xi - 237{\xi}^2 + 124{\xi}^3 - 
          15{\xi}^4 \right) }{1200}  
\right.\non\\&+& \left.
     \frac{\rho^4\left( -438 + 766\xi - 337{\xi}^2 + 8{\xi}^3 + 
          25{\xi}^4 \right) }{48{\left( 1 - \xi \right) }^2} \right] 
   L_\xi
\eea

\end{document}